\documentstyle[11pt]{article}
\def\be{\begin{equation}}
\def\bea{\begin{array}}
\def\eea{\end{array}}

\def\ka{\kappa}
\def\ee{\end{equation}}
\def\ep{\epsilon}
\begin{document}
\begin{center}
\LARGE {\bf Generating Functional and Large N-Limit of  Nonlocal 2D
Generalized Yang-Mills Theories ($nlgYM_2$'s)}
\\
\end{center}
\vskip 1.5cm
{\centerline {\bf K. Saaidi \footnote {e-mail: lkhled@molavi.ut.ac.ir},
 \bf H. M. sadjadi  \footnote {e-mail:  amohosen@khayam.ut.ac.ir}}}
   \vskip 0.2cm
\begin{center}
{\it  Department of Physics, University of Tehran,
North Karegar St., Tehran, Iran. }\\
\vskip 3cm
{\bf{Abstract}}\\
\end{center}

Using the path integral method, we calculate the partition function and 
generating functional (of the field strengths) on the nonlocal generalized 
$2D$ Yang - Mills theories ($nlgYM_2$'s), which  is 
nonlocal in auxiliary field [14]. Our calculations is done for general surfaces.
We find a general expression for free energy of $W(\phi ) = \phi^{2k}$ in
$nlgYM_2$ theories at the strong coupling phase (SCP) regime ($A > A_c$) 
for large groups. In the specific $\phi^4$ model, we show that the 
theory has a third order phase transition.

\newpage
{\section{Introduction}}

This paper will be devoted to a renewed study of two dimensional Yang-Mills
theory without matter, a system which can be easily solved. Yet we will see
that there is still much to say about this system.  Pure two dimensional
Yang-Mills theories($YM_2$'s) have certain properties, such as
invariance under area preserving diffeomorphism and lack of any propagating
degrees of freedom[1]. There are, however, ways to generalize these theories 
with out losing those properties. One way so - called
 generalized Yang-Mills theories($gYM_2$'s) [2] is
\be
iTr(B\epsilon^{\mu \nu}F_{\mu \nu}) + f(B)
\ee

Here $F_{\mu \nu}$ is the Yang-Mills field strength and B is a
scalar field in the adjoint representation of the gauge group. Standard
dimensional analysis applied to (1) gives $F_{\mu \nu}$ dimension 2 and B 
dimension 0, so power counting allows an arbitrary class function $\it f(B)$.
This model produce by E. Witten [2] and has obtained the partition
function by considering its action as a perturbation of the topological
theory at zero area. In [3-5] the Green function,  partition function
and  expectation values of Wilson loops were calculated. One can, however, 
use standard path - integration and calculate the observables of the theory
[6,7]. To study the behaviour of these theories for large groups is also interest.
This was studied in [8-11] for ordinary $YM_2$ theories and in [12,13] for
$gYM_2$ theories. It was shown that $YM_2$'s and some classes of $gYM_2$'s
have a third - order phase transition in a certain area. There is another way
to generalize $YM_2$ and $gYM_2$, and that is to use a non - local action for 
the auxiliary field, which so -  called nonlocal $YM_2 (nlYM_2$'s) and nonlocal 
$gYM_2 (nlgYM_2$'s) theories, respectively [14]. The authors of [14] studied 
$nlYM_2$ and investigated the order of transition for that. We want study 
the wave  function, partition function, generating functional of $nlgYM_2$ 
and also their properties for large gauge group in the state which
 $W(\phi) = \phi^4$. The scheme of the present paper is the following.

In sec.2, the wave function and partition function of $nlgYM_2$  on general 
surfaces are computed. 
In sec.3, the generating functional of $nlgYM_2$ on disk and general surfaces
are calculated. In sec.4, the properties of $nlgYM_2$ large groups, for the 
case which $f(B) = Tr(B^{2k})$, are studied. Finally in sec.5, we test
our theory for $\phi^4$ model $(f(B) = Tr(B^4))$. It is shown that the large 
group properties of $nlgYM_2$ are the same which was found for ordinary $gYM_2$.

{\section{The Wave Function of $nlgYM_2$}}

The $nlgYM_2$ is defined by [14]
\be
e^S := \int DB exp{\Biggr \{}i\int Tr(BF)d\mu + \omega{\Biggr [}\int f(B)
d\mu{\Biggl ]}{\Biggl \}},
\ee
where $d\mu$ is the invariant measure of the surface 
\be
d\mu :=\frac{1}{2}\ep_{\mu \nu}dx^{\mu}dx^{\nu}.
\ee
F is the field strength corresponding to the gauge field and B is a 
pseudo - scalar field in the adjoint representation of the group.  Along 
the line of [7,14], we begin by calculating the wave function on a disk. 
we obtain
\be 
\psi_D(U) = \int DFe^S \delta {\Biggr (}Pexp\oint_{\partial D}A , U{\Biggl )}.
\ee
Here U is the class of the Wilson loop corresponding to the boundary.
The delta function is also a class delta function, in which, its support
the boundary conditions. This delta function can be expanded in terms of the 
characters of irreducible unitary representations of the group; i.e.
\be
\delta {\Biggr (}Pexp\oint_{\partial D}A , U{\Biggl )} = 
\sum_R \chi_R(U^{-1})\chi_R{\Biggr (}Pexp\oint_{\partial D}A {\Biggl )}.
\ee
We introduce Fermionic variables $\eta$ and ${\bar \eta}$ in the 
representation R to write the Wilson loop as [6,7]     
\be
\chi_R{\Biggr (}Pexp\oint_{\partial D}A{\Biggl )} = 
\int D\eta D{\bar \eta}exp
{\Biggr \{} \int_0^1 dt{\bar \eta}(t){\dot \eta}(t) + 
\oint_{\partial D}{\bar \eta}A\eta{\Biggl \}}
\eta^{\alpha}(0){\bar \eta}_{\alpha}(1). 
\ee

Inserting (6) in (5) and then (4), using the Schwinger - Fock gauge, and 
integrating over, F, B, and the Fermionic variables, respectively, one 
obtains   
\be
\psi_D(U) = \sum_R \chi_R(U^{-1})d_R exp{\{ \omega [AC_f(R)] \}},
\ee

Here $d_R$ is the dimension of the representation R and 
\be 
C_f(R)1_R =: f(-iT_R). 
\ee
and $f(-iT_R)$ means that one has put $-iT^a$ in the representation R 
instead of $B^a$ in the function ${\it f}$.
Where as the action of the original B-F theory (2) is not extensive; i.e. 
\be
S_{A_1+A_2}(B, F) \neq S_{A_1}(B, F) + S_{A_2}(B, F).
\ee

Therefore, one  cannot simply glue the disk wave function to obtain, the 
wave function corresponding to a larger disk. To obtain the wave function 
for an arbitrary surface, however, one can begin with a disk of the same  
area and impose boundary conditions on certain parts of the boundary of the 
disk. These conditions are those corresponding to the identifications 
needed for constructing the desired surface from a disk. The only things 
to be  calculated are integrations over group of characters of the same
representation [5]. This is easily done and one arrives at
\be
\psi_{\sum_{g,q}}(U_1,\ldots, U_n) = \sum_R h_R^q d_R^{2-2g-q-n} 
\chi_R(U_n^{-1}) \ldots \chi_R(U_n^{-1}) exp{\{\omega[-C_f(R)
A_{\sum_{g,q}}]\}},
\ee
where $\sum_{g,q}$ is a surface containing ${\it g}$ handles, ${\it n}$ 
boundaries and ${\it q}$ projective planes. $h_R$ is defined as 
\be
h_R := \int dU \chi_R(U^2),
\ee
$h_R = 0 $ unless the representation R is self conjugate. In this case,  
this representation has an invariant bilinear form. Then, $h_R=1$ if this 
form is symmetric and $h_R=-1$ if it is antisymmetric[15].

The partition function of the theory on a sphere is obtained if we put $U_i$'s
equal to unity and ${\it g}$ and ${\it q}$ equal to zero. We obtain
\be
Z_{s^2} = \sum_R d_R^2 exp{\{\omega[-AC_f(R)]\}}.
\ee

{\section{The Generating Functional Z[J] of $nlgYM_2$}}

To calculate the Green functions of the strength $F^a$'s, we again begin with
the disk and calculate the wave function of $nlgYM_2$ on the disk, with a 
source term coupled to F; i.e.
\be 
\psi_D[J] = \int DFe^{\{S + \int Tr(FJ)d\mu \}} 
\delta {\Biggr (}Pexp\oint_{\partial D}A , U{\Biggl )}.
\ee

Following the same steps of the previous section, we arrive at
\be
\psi_D[J] = \sum_R \chi_R(U^{-1})Tr_R{\Biggr \{}Pexp{\Biggr (}\omega{\Biggr [}
\int f(iJ^a(x) + iT^a)d\mu{\Biggl ]}{\Biggl )}{\Biggl \}}.
\ee

In the above equation P stands for ordering according to the angle variable
on the disk. to obtain the generating functional Z[J] of $nlgYM_2$ for an 
arbitrary surface, $\sum_{g,q}$, we can use the same procedure which was
used in obtaining (10) and the result is 
\be
Z_{\sum_{g,q}}[J]=\sum_R h_R^q d_R^{2-2g-q-1}exp{\{\omega[AC_f(R)]\}}
Tr_R{\Biggr \{}Pexp{\Biggr (}\omega {\Biggr [}\int f(iJ^a+iT^a)d\mu{\Biggl ]}
{\Biggl )}{\Biggl \}}.
\ee

As an example, consider $YM_2$, in which $\omega[\int f(B)d\mu]=-\frac{1}{2} 
\ep \int Tr(B^2)d\mu$. In this case (15) reduces to 
\be
Z_{\sum_{g,q}}[J]=Z_1[J]
\sum_R h_R^q d_R^{2-2g-q-1}exp\{-\frac{\ep}{2}C_2(R)A_{\sum_{g,q}}\}
Tr_R{\Biggr \{}Pexp{\Biggr (} \ep \int dt \int ds \sqrt{g} J(t,s)
{\Biggl )}{\Biggl \}},
\ee
where
$$Z_1[J] = exp{\Biggr (}-\frac{\ep}{2}\int J^aJ_a d\mu{\Biggl )}.$$

Which is in agreement with the result obtain in [7]. Functional 
differentiating of (15) with respect to $J(x)$ gives us the $n$-point 
functions of $F$'s in the Schwinger-Fock gauge. 

{\section{Large N-Limit of $nlgYM_2$}}

Starting from (12), consider the case that gauge group is $U(N)$. The 
representation of this group are labeled by $N$ integers $n_i$ satisfying
\be
 n_i \geq n_j ,\hspace{2cm} i \leq j .
 \ee

 The dimension of this representation is
\begin{equation}
d_R= \prod_{1 \leq i \leq j \leq N} (1+\frac{n_i-n_j}{j-i}),
\end{equation}
and the $k-th$ Casimir is 
\begin{equation}
C_k(R) = \sum_{i=1}^{N}[(n_i+N-i)^k - (N-i)^k].
\end{equation}

Taking $C_f(R)$ a linear function of the Casimirs (19) and redefine the 
function $\omega$ and introduce another function as
\be
-N^2V[A\sum_{k=1}^N a_k {\hat C_k}(R)] := \omega [-A C_f(R)],
\ee
where
\be
{\hat C_k}(R) = \frac{1}{N^{k+1}}\sum_{i=1}^N (n_i+N-i)^k.
\ee

Then, following [10], we use the definitions 
\be
x:=\frac{i}{N},
\ee
and
\begin{equation}
\phi(x) =\frac{i-n_i-N}{N}.
\end{equation}

So apart from an unimportant constant, the partition function takes the form
\begin{equation}
Z[\phi (x)] = \int D\phi (x) e^{\{-N^2S(\phi)\}},
\end{equation}
where
\begin{equation}
S(\phi ) = V{\Biggr (}A \int _{0}^{1} W[\phi (x)] dx{\Biggl )}
+ \int_{0}^{1} dx \int_0^1 dy log|\phi (x)- \phi (y)|, 
\end{equation}
and
\be
   W(\phi ) := \sum_{k=1} (-1)^k a_k \phi^k.
\ee

In the large N-limit, only the configuration of $\phi$ contributes to the
partition function that minimizes $S$. To find it, we puts variation of $S$ 
with respect to $\phi$ equal to zero.

\begin{equation}
\frac{{\hat A}}{2}W'(\phi ) =
P\int_0^1 \frac{dt}{\phi (x) -\phi (x)},
\end{equation}
where  
\be
{\hat A} := AV'{\Biggr [}A\int_0^1 dx W(\phi (x)){\Biggl ]}.
\ee

One defines a density function for $\phi$ as 
\be
u(\phi ) := \frac{dx(\phi )}{d\phi }|_{\phi=z},
\ee
which should be positive and normalized to
\be
\int_{-a}^{a} u(z) dz = 1.
\ee

Then (27) becomes
\begin{equation}
\frac{\hat A}{2}W'(z) = P\int_{-a}^a \frac{u(t)dt}{z - t}.
\end{equation}
To solve (31), we defined the function $H(z)$ on the complex z-plane [10]
\be
H(z) := \int_{-a}^a\frac{u(t)dt}{z-t}.
\ee

This function is analytic on the complex plane, except for a cut at
$[-a, a]$. With proceed the same procedure which was followed in [12], 
one arrives at
\be
H(z)= \frac{{\hat A}}{2}W'(z) - \sqrt{z^2-a^2}\sum_{m,n=0}^{\infty}M_n 
\frac{a^{2n}z^m}{(2n+m+1)!}g^{(2n+m+1)}(0),
\ee
where
\be
g(z) = \frac{{\hat A}}{2} W'(z),
\ee
and
\be
M_n = \frac{(2n-1)!!}{2^nn!}, \hspace{2cm} M_0=1.
\ee

$g^{(k)}$ is the $k-th$ derivative of $g$ with respect to z.
From (32), it is seen that
\be
ImH(z+i \ep )= -\pi u(z), \hspace{2cm} x \in [-a, a]
\ee
which gives
\be
u(z) = \frac{\sqrt{a^2-z^2}}{\pi} \sum_{n,m=0}^{\infty}
\frac{M_na^{2n}z^mg^{(2n+m+1)}(0)}{(2n+m+1)!}.
\ee

To obtain $a$, one can use (30) and (37), which yields
\be
\sum_{n=0}^{\infty} \frac{M_na^{2n}g^{(2n-1)}(0)}{(2n-1)!}=1.
\ee

Defining a free energy  function as
\be
F := -\frac{1}{N^2}S|_{\phi_{cla.}}
\ee

It is seen that
\be
F'(A)= V'(A\kappa) \kappa,
\ee
where
\be
\kappa = \int_0^1 W[\phi(x)]dx = \int_{-a}^a u(z)W(z)dz.
\ee

By making use of equations (37) and an explicit expression for $W(z)$
as a function of $\it z$, we can calculate $\kappa$ and therefore
at last we compute $F'_w(A)$ (40) for this model. Note that
the above solution is valid in the weak $(A \leq A_c)$ regime, where $A_c$
is the critical area. If $A > A_c$, then the constraint $u \leq 1$ is 
violated.

{\section{The $W(z) = z^{2k}$ Model for $nlgYM_2$}}

{\subsection{ WCP Regime($A\leq A_c$)  }}

In order to study the behaviour of any model in the SCP
regime $( A > A_c)$, we need to know  the explicit form of density function
in the weak regime, $u_w(z)$. So by rewriting (37), (38) and (40) 
for $z^{2k}$ model, one can arrives at
\be
u_w(z) = \frac{k{\hat A}}{\pi}\sqrt{a^2-z^2}
\sum_{n=0}^{k-1}M_na^{2n}z^{2k-2n-2},
\ee
\be
\frac{k{\hat A}a^{2k}}{2^k}Q(k) = 1,
\ee
\be
F'_w(A) = \frac{kV'{\hat A}a^{4k}}{2^k}E(k),
\ee
where
$$ Q(k) = \sum_{n=0}^{k-1}\frac{(2k-2n-3)!!(2n-1)!!}{(k-n-1)!(n+1)!}$$
\be
E(k) = \sum_{n=0}^{k-1}\frac{(2k-2n-3)!!(2k+2n-1)!!}{(k-n-1)!(k+n+1)!}
\ee

This is, of course, in complete accordance with [13]. But one must now obtain 
the quantities in terms of A not ${\hat A}$. It is seen that 
\be
 F'_w(A) = \frac{E(k)}{kAQ^2(k)} = \frac{1}{2kA}.
\ee

The function V is disappeared from $F'_w(A)$, as it can be seen by the 
rescaling ${\hat \phi}:= A^{\frac{1}{2k}}\phi$.\\
This completes our discussion of the weak-region $nlgYM_2$. As A increases, 
a situation is encountered where $u_w$ exceeds 1. This density function
is, however, not acceptable, as it violate the condition (17).

{\subsection{SCP Regime $(A>A_c)$}}

One of the interesting point of the $Z^{2k}(k>1)$ model is the fact which the 
density function in weak-region (42) has only one minimum at $z=0$, and 
two maxima which are symmetric with respect to origin [13]. So that to find the 
density function in strong-region, will be relevant with three cut Cauchy 
problem. Hence following [12], we use the following ansatz for $u_s$
\be
u_s(z)=\left\{ \bea{cc}
{\hat u_s(z)} \;\;\;\; z\in L := [-a , -b] \cup [-c, c] \cup [b , a]\\
1 \;\;\;\;\;\;\;\;\;\;\;\; z\in L':=[-b ,-c] \cup [c, b]
\eea\right.
\ee

Using methods exactly the same as those used in [12], one must solve
\begin{equation}
\frac{{\hat A}}{2}W'(z) = P\int_{-a}^{a} \frac{u_s(t)dt}{z-t}   \;\;\;\;\;,\;\;\;\; z \in L,
\ee
and
\begin{equation}
\int_{c}^{b}{\Biggr \{}\frac{{\hat A}}{2}W'(z) - P\int_{-a}^{a} 
\frac{u_s(t)dt}{z-t}{\Biggl \}}dz = 0.
\ee

To do so, one defines a function $H_s$ as 
\be
H_s(z)= \int_{-a}^a \frac{u_s(t)dt}{z-t},
\ee
which is found to be
\be 
H_s(z) =  k{\hat A}z^{2k-1} + 2T(z){\Biggr [}k\frac{{\hat A}}{2}\sum_{[n_i]=0}^{\prime}
\tau(n_1,n_2,n_3)z^{2n_4} - \int_c^b \frac{tdt}{(z^2-t^2)T(t)}{\Biggl ]} ,
\ee
where the prime on the $\sum$ indicate the following condition
\be
\sum_{i=1}^4 n_i = k-2,
\ee
and
\be
T(z) = \sqrt{(a^2-z^2)(b^2-z^2)(c^2-z^2)},
\ee
\be
\tau (n_1,n_2,n_3) = M_{n_1}M_{n_2}M_{n_3}a^{2n_1}b^{2n_2}c^{2n_3}.
\ee

Using the fact that $H_s(z)/T(z)$ should behave as $\frac{1}{z^4}$ for 
large $z$, one obtains
\be 
k{\hat A}\sum_{[n_i]=0}^{\prime}\tau (n_1,n_2,n_3) = 2 \int_c^b 
\frac{tdt}{T(t)},
\ee
\be
k{\hat A}\sum_{[n_i]=0}^{\prime}\tau (n_1,n_2,n_3) =  1 + 2 \int_c^b 
\frac{t^3dt}{T(t)}.
\ee

Where the prime over summations in (55) and (56) indicates the 
following conditions, respectively
\be 
\sum_{i=1}^3 n_i = k-1,
\ee
\be
\sum_{i=1}^3 n_i = k.
\ee

In order to obtain the parameters a, b and c in spite of (55) and (56) we need 
another equation (50) which is found by expresses the action in terms of 
$u_s(z)$ and minimize that along with the (30), as a constraint [11,12]. 
By expanding (50) and (51) at large z and compare them, one 
can easily arrives at
\be
F'_s(A)= V'_s(A\kappa_s){\Biggr \{}k{\hat A} \sum_{[n_i]=0}^{\prime}
\tau(n_1,n_2,n_3)\tau_1(n_4,n_5,n_6)
 + 2\sum_{[n_i]=0}^{\prime}\tau_1(n_1,n_2,n_3)
 \int_c^b \frac{t^{2n_4+1}dt}{T(t)}{\Biggl \}},
 \ee
where the prime over first and second summation indicate the following 
constraint, respectively
\be
\sum_{i=1}^6 n_i = 2k,
\ee
\be
\sum_{i=1}^4 n_i = k+1,
\ee
and
\be
\tau_1(n_1,n_2,n_3) = \frac{a^{2n_1}b^{2n_2}c^{2n_3}}{2^{n_1+n_2+n_3}} 
\prod_{i=1}^3 \frac{(2n_i-3)!!}{n_i!},
\ee
where, we define $(-3)!! = -1.$

Equation (59), is an explicit relation for $F'_s(A)$, which represents the  
SCP regime of our theory. It is seen that the structure of $F'_s(A)$
is very complicate, therefore, as example, we can study the order of 
transition for $z^4$ model(k=2).
 
{\section{The $z^4$ Model of  $nlgYM_2$}}
 
{ \subsection{WCP Regime  $(A \leq A_c)$}}

 In the previous section we study the $nlgYM_2$ for $ z^{2k}$ model. 
 In this section we can check the result of it for $z^4$ model. By 
 rewriting eqs.(42-45), we have
 \be
 u_w(z) = \frac{{\hat A}}{\pi}\sqrt{a^2-z^2}(a^2 + 2z^2),
 \ee
 \be
 \kappa_w = \frac{3a^4}{16}
 \ee
 \be
{\hat A} = \frac{4}{3a^4}
 \ee
 and
 \be
 F'_w(A) = \frac{1}{4A}
 \ee
 It is see that, the density function in WCP regime, $u_w(z)$, has a
 minimum at $\it z=0$, and two maxima at ${\it z}_{1,2}= \pm{\frac{a}
 {\sqrt 2}} $. Equations (63-66) are valid in the regime which $a \leq a_c =
 \frac{8}{3\sqrt{2} \pi}$ or $A \leq A_c$. 
 The value of $A_c$, is obtained from
 \be
 u_w(z_{1,2}) = 1,
 \ee
 which gives
 \be
 A_cV'_{c}{\Biggr (}\frac{32A_c}{27\pi^4}{\Biggl )} = \frac{27 \pi^4}{256}.
 \ee

In spite of some constant, these almost are the same results which have 
been calculated for local $gYM_2$ theory [12]. 

{\subsection{SCP Regime ($A>A_c$)}}

The $z^4(z)$ model for $nlgYM_2$, is a state which, the density 
function in WCP have a minimum at origin and two maxima which are symmetric 
with respect to origin. So one can use of the results in previous section 
and arrives at
\be
\int_c^b {\Biggr \{}2{\hat A}z^3 - P\int_{-a}^a \frac{u_s(t)dt}{z-t}
{\Biggl \}}dz = 0,
\ee
\be
{\hat A}(a^2+b^2+c^2) = 2\int_c^b\frac{tdt}{T(t)},
\ee
\be
{\hat A}{\Biggr \{}(a^2b^2+a^2c^2+b^2c^2)+\frac{3}{2}(a^4+b^4+c^4){\Biggl \}} 
= 2 + 4\int_c^b\frac{t^3dt}{T(t)}.
\ee

Finally by making use of eqs.(54,59,62), it is seen that  

$$F'_s(A) = V'(A\ka_s){\Biggr \{}\frac{{\hat A}}{16}{\Biggr [}
\frac{5}{4}(a^8+b^8+c^8)
-\frac{1}{2}(a^4b^4+a^4c^4+b^4c^4)+(a^2b^2c^4+a^2c^2b^4+a^4b^2c^2)$$
$$-(a^2b^6+a^2c^6+b^2a^6+b^2c^6+c^2a^6+c^2b^6){\Biggl ]}$$
$$+\frac{1}{8}{\Biggr [}a^6+b^6+c^6-(a^2b^4+a^2c^4+b^2a^4+b^2c^4+c^2a^4+
c^2b^4)+ 2a^2b^2c^2{\Biggl ]}\int_c^b\frac{tdt}{T(t)}+$$
\be
\frac{1}{4}{\Biggr [}a^4+b^4+c^4-2(a^2b^2+a^2c^2+b^2c^2){\Biggl ]}\int_c^b
\frac{t^3dt}{T(t)}+(a^2+b^2+c^2)\int_c^b \frac{t^5dt}{T(t)}-
2\int_c^b \frac{t^7dt}{T(t)}{\Biggl \}}
\ee

By using the same procedure was used in [12,13] and expand eqs.(69-72) near 
the critical point and  then solve them with together, we can obtain
\be
F'_s(A) = \frac{V'_s}{{\hat A}}[\frac{1}{4} + \frac{\beta}{27}{\alpha}^2+
\ldots],
\ee
or
\be
F'_s(A) - F'_w(A) = \frac{\beta}{27A_c}{\alpha}^2 + \ldots,
\ee
where 
\be
{ \alpha} =(\frac{A - A_c}{ A_c})^2,
\ee
and
\be 
\beta = {{\Biggr (} 1 + \frac{A_c \kappa_{cs}V''_{cs}}{V'_{cs}}{\Biggl )}}^2.
\ee
It is seen that the theory, for $\phi^4$ model has a third order phase 
transition, which is in agreement  with ordinary $gYM_2$.

\end{document}